\documentclass{eas}
\usepackage{graphicx}
\runningtitle{Nova eruptions}
\begin{document}

\title{Nova Eruptions with Infrared \\ Interferometric Observations} 
\author{Val\'erio A. R. M. Ribeiro}\address{Department of Astrophysics/IMAPP, Radboud University, PO Box 9010, NL-6500 GL Nijmegen, The Netherlands}
\begin{abstract}
Infrared interferometric observations have a great deal of potential to unravel the nature of the nova eruptions. We suggest that techniques, already in place, to derive the ejection details at optical wavelengths be used with infrared interferometric observations to derive parameters such as the ejected mass in a nova eruption. This is achievable based on modelling the initial phase of the eruption when the infrared light is dominated by the free-free thermal process.
\end{abstract}
\maketitle
\section{Introduction}
Classical and Recurrent Novae -- distinguished by the observation of one or more eruptive events, respectively -- are thermonuclear runaways (TNR) on the surface of a white dwarf (WD) star which ensues following extensive accretion of hydrogen rich material from a less evolved secondary star. The ejected material is rich in heavy elements which enrich the interstellar medium (\cite{GTW98,BE08,GEW14}, see also Gehrz et al., these proceedings).

The nova eruptions are the most common thermonuclear runaways in the Galaxy (34$^{+15}_{-12}$ novae per year; \cite{DBK06,S97}) and scale with galaxy size (\cite{SCP14}). The infrared temporal evolution of the nova is nicely described in Gehrz et al. (these proceedings) and for detailed reviews on interefometric observations of novae, see Chesneau \& Banerjee (2012) and Chesneau (2014). Of particular interest, to this work, is that during the initial phases of the eruption the ejected material can be modelled as an optically thick fireball that radiates like a blackbody and as the expansion continues and reduces in density, the infrared continuum becomes dominated by optically thin free-free emission with strong hydrogen re-combination lines on top.

Observations of novae are going through a renaissance over the last decade or so driven by the upgrade of telescopes, instruments, and new discoveries (see the various contributions in \cite{WR14}). Of particular note, the discovery of high energy $\gamma$-ray emission (\cite{AAA10,AAA14}) and pinpointing the location of $\gamma$-ray production, from internal shocks due to different velocity components at the equatorial and polar regions (\cite{CLY14}).

A number of factors are required to be systematically studied across the electromagnetic spectrum in order to understand the evolution of the nova eruption. Of paramount importance, in any field of astrophysics, is the distance to an object. In the Milky Way, the distance to novae allow us to derive their spatial distribution in order to understand from which population/s novae arise from -- disc/bulge -- and also the occurrence of novae in our Galaxy.

\section{Infrared Interferometric Observations of Novae in Eruption}
The first observations with an infrared interferometer occurred with nova V1974~Cygni (1992) on the Mk III Interferometer (\cite{QEM93}). The nova was resolved aroudn 10 days after maximum light. Fitting a uniform disk model to the data the distance was estimated to be 2.5 kpc, in line with other studies (\cite{PLH95}). There was one more nova-like eruptive event, in V838 Mon (\cite{LRT05}), that was followed with Palomar Testbed Interferometer (PTI). However, this event is largely accepted to have been the merger of two stars.

It was not until the recurrent nova RS Ophiuchi erupted in 2006 that infrared interferometric observations truly come to play. RS Oph was observed first with the Infrared Optical Telescope Array, Keck and PTI (\cite{MBT06}), suggesting that the near-infrared emission arose due to a non-expanding, dense, and ionised circumbinary gaseous disk or reservoir. However, Lane et al. (2007) resolve the emission in RS Oph using PTI providing clear evidence for a near-infrared source that initially expanded and then began shrieking.

The AMBER instrument on the Very Large Telescope Interferometer (VLTI) was also used, on day 5.5 after eruption, to measure the $K$-band continuum and the Br$\gamma$ and HeI 2.06 $\mu$m lines (\cite{CNM07}). The $K$-band emission is dominated by free-free emission and had a smaller size than the Br$\gamma$ and HeI lines (3x2 mas, 5x3 mas, and 6x4 mas, respectively). These results were also contrary to Monnier et al. (2006). Furthermore, Chesneau et al. (2007) found two velocity fields in the Br$\gamma$ line; a slowly expanding ring-like structure ($v_{\rm rad} \leq$~1800~km~s$^{-1}$), and a fast structure extended in the E-W direction ($v_{\rm rad} \sim$~2500--3000~km~s$^{-1}$). A two velocity component was also required to replicate the Hubble Space Telescope narrow band imaging observations of the resolved remnant, at day 155 after eruption, and the ground-based optical spectroscopy (\cite{RBD09}). Lastly, RS Oph observations with the Keck Interferometer Nuller, around day 3.8 after eruption, showed evidence for dust that is present in-between eruptions, rather than created during the eruption (\cite{BDT08}) in line with findings from the Spitzer Space Telescope of silicate dust that survives the hard radiation impulse and shock blast wave from the eruption (\cite{EWH07}).

Other novae have been observed since RS Oph with infrared interferometric instruments; the recurrent nova T~Pyxidis (\cite{CMB11}), the dust forming nova V1280 Sco (\cite{CBM08,CLO12}), and the $\gamma$-ray detected classical nova V339 Delphini (2013; \cite{SBG14}).

\section{Prospects for the Future}
What is clear from the various bodies of work is that the nova eruption is far from spherical, or uniform disks. Efforts are now under way to understand the progress of the nova eruption from optical to radio observations applying one single model. For example, optical emission line profile fitting of the classical nova V959~Mon proved useful in determining, the ejection morphology, inclination angle and the expansion velocity of the ejecta (\cite{RMV13,SGS13}). These results are also corroborated with the observations of eclipses at x-ray and optical wavelengths (\cite{POW13,MDC13}) and observations of a bipolar ejection morphology at radio frequencies (\cite{CLY14,LRC15}). Lastly, applying the free-free thermal process to the models derived at optical wavelengths above we were able to derive the distance to V959~Mon (\cite{LRC15}).

Up to now the observations have provided important information on the distances to these objects assuming uniform/Gaussian fitting to the visibilities/images when available as well as departure from these morphologies for the ejecta. In fact, the distance may also be constrained assuming the Blackbody angular radius and Doppler expansion velocity which give day of the eruption, the distance and the outburst luminosity (\cite{EG12} and Gerzh et al. these proceedings). However, we can extract further information if we assume a thermal free-free emission. Here we will, also, be able to extract ejected masses, temperature of the ejecta, and density profiles of the ejecta. Ideally, we would want to combine all this information with other wavelengths, such as optical for the Doppler expansion velocity, in order to reduce the amount of free parameters and computational time. Although the distance is a very important issue, the GAIA satellite will largely solve this in the coming years. Furthermore, a suit of new instruments are to be installed in the VLTI which will provide marked improvement, both in resolution and sensitivity (MATISSE and GRAVITY), over current capabilities.

\acknowledgements
The author would like to thank Bob Gehrz for reading and commenting on the manuscript and Christiaan Brinkerink for useful discussion on interferometric observations. The author also acknowledges financial support from the Radboud Excellence Initiative.

\section*{Questions}
{\bf Q: Lizette Guzman:} Have you used the expansion velocity in the optical compared to the radio emission?

{\bf A: Val\'erio Ribeiro:} Yes. We see this, at least in one case that comes to mind, when comparing the optical and radio imaging in RS Oph (\cite{RBD09,OBP06}).

{\bf Q: Jose Groh:} I had the impression that your models have less flux at the central regions than observed at later times. Could you comment on the reasons for this behaviour and whether this is significant in the context of your models?

{\bf A: Val\'erio Ribeiro:} The models I showed assumed free-free thermal emission, so as input to the models we have ejection velocity, ejected mass, and temperature. We did not have time to produce a full set of parameter space in order to find the best fit. So what you are seeing is that the models with chosen input parameters became optically thin quicker than the observations. We really need to explore the fuller parameter space.


\begin{thebibliography}{99}
\bibitem[Abdo et al. 2010]{AAA10} Abdo, A.~A., Ackermann, M., Ajello, M., et al.\ 2010, Science, 329, 817 
\bibitem[Ackermann et al. 2014]{AAA14} Ackermann, M., Ajello, M., Albert, A., et al.\ 2014, Science, 345, 554
\bibitem[Barry et al. 2008]{BDT08} Barry, R.~K., Danchi, W.~C., Traub, W.~A., et al.\ 2008, ApJ, 677, 1253 
\bibitem[Bode \& Evans 2008]{BE08} Bode, M. F. \& Evans, A.\ 2008, editors {\em Classical Novae}, 2nd Edition (Cambridge University Press)
\bibitem[Chesneau 2014]{C14} Chesneau, O.\ 2014, in {\em Stella Novae: Past and Future Decades}, eds. P. A. Woudt \& V. A. R. M. Ribeiro, ASPCS, Vol. 490, 243
\bibitem[Chesneau \& Banerjee 2012]{CB12} Chesneau, O. \& Banerjee, D.~P.~K.\ 2012, BASI, 40, 267 
\bibitem[Chesneau et al. 2008]{CBM08} Chesneau O., Banerjee D. P. K., Millour F., et al.\ 2008, A\&A, 487, 223
\bibitem[Chesneau et al. 2009]{CCL09} Chesneau O., Clayton G. C., Lykou F., et al.\ 2009, A\&A, 493, L17
\bibitem[Chesneau et al. 2012]{CLO12} Chesneau, O., Lagadec, E., Otulakowska-Hypka, M., et al.\ 2012, A\&A, 545, A63 
\bibitem[Chesneau et al. 2011]{CMB11} Chesneau O., Meilland A., Banerjee D. P. K., et al.\ 2011, A\&A, 534, L11
\bibitem[Chesneau et al. 2007]{CNM07} Chesneau O., Nardetto N., Millour F., et al.\ 2007, A\&A, 464, 119
\bibitem[Chomiuk et al. 2014]{CLY14} Chomiuk, L., Linford, J.~D., Yang, J., et al.\ 2014, Nature, 514, 339 
\bibitem[Darnley et al. 2006]{DBK06} Darnley, M.~J., Bode, M.~F., Kerins, E., et al.\ 2006, MNRAS, 369, 257
\bibitem[Evans \& Gehrz 2012]{EG12} Evans, A., \& Gehrz, R.~D.\ 2012, Bulletin of the Astronomical Society of India, 40, 213 
\bibitem[Evans et al. 2007]{EWH07} Evans, A., Woodward, C.~E., Helton, L.~A., et al.\ 2007, ApJL, 671, L157 
\bibitem[Gehrz et al. 2014]{GEW14} Gehrz, R. D., Evans, A., Woodward, C. E.\ 2014, in {\em Stella Novae, Past and Future Decades}, eds. P. A. Woudt \& V. A. R. M. Ribeiro, ASPCS, Vol. 490, 227
\bibitem[Gehrz et al. 1998]{GTW98} Gehrz, R.~D., Truran, J.~W., Williams, R.~E., \& Starrfield, S.\ 1998, PASP, 110, 3
\bibitem[Lane et al. 2005]{LRT05} Lane, B.~F., Retter, A., Thompson, R.~R., \& Eisner, J.~A.\ 2005, ApJL, 622, L137 
\bibitem[Lane et al. 2007]{LSB07} Lane, B.~F., Sokoloski, J.~L., Barry, R.~K., et al.\ 2007, ApJ, 658, 520 
\bibitem[Linford et al. 2015]{LRC15} Linford, J.~D., Ribeiro, V.~A.~R.~M., Chomiuk, L., et al.\ 2015, ApJ, 805, 136 
\bibitem[Monnier et al. 2006]{MBT06} Monnier, J.~D., Barry, R.~K., Traub, W.~A., et al.\ 2006, ApJL, 647, L127 
\bibitem[Munari et al. 2013]{MDC13} Munari, U., Dallaporta, S., Castellani, F., et al.\ 2013, MNRAS, 435, 771 
\bibitem[O'Brien et al. 2006]{OBP06} O'Brien, T.~J., Bode, M.~F., Porcas, R.~W., et al.\ 2006, Nature, 442, 279 
\bibitem[Page et al. 2013]{POW13} Page, K.~L., Osborne, J.~P., Wagner, R.~M., et al.\ 2013, ApJL, 768, L26 
\bibitem[Paresce et al. 1995]{PLH95} Paresce, F., Livio, M., Hack, W., \& Korista, K.\ 1995, A\&A, 299, 823 
\bibitem[Quirrenbach et al. 1993]{QEM93} Quirrenbach, A., Elias, N.~M., II, Mozurkewich, D., et al.\ 1993, AJ, 106, 1118 
\bibitem[Ribeiro et al. 2009]{RBD09} Ribeiro, V.~A.~R.~M., Bode, M.~F., Darnley, M.~J., et al.\ 2009, ApJ, 703, 1955 
\bibitem[Ribeiro et al. 2013]{RMV13} Ribeiro, V.~A.~R.~M., Munari, U., \& Valisa, P.\ 2013, ApJ, 768, 49 
\bibitem[Schaefer et al. 2014]{SBG14} Schaefer, G.~H., Brummelaar, T.~T., Gies, D.~R., et al.\ 2014, Nature, 515, 234 
\bibitem[Shafter 1997]{S97} Shafter, A.~W.\ 1997, ApJ, 487, 226
\bibitem[Shafter et al. 2014]{SCP14} Shafter, A.~W., Curtin, C., Pritchet, C.~J., Bode, M.~F., \& Darnley, M.~J.\ 2014, in {\em Stella Novae: Past and Future Decades}, eds. P. A. Woudt \& V. A. R. M. Ribeiro, ASPCS, Vol. 490, 77 
\bibitem[Shore et al. 2013]{SGS13} Shore, S.~N., De Gennaro Aquino, I., Schwarz, G.~J., et al.\ 2013, A\&A, 553, A123 
\bibitem[Woudt \& Ribeiro 2014]{WR14} Woudt, P.~A. \& Ribeiro, V. A. R. M.\ 2014, editors {\em Stella Novae, Past and Future Decades}, ASPCS, Vol. 490
\end{thebibliography}
\end{document}